\documentclass[fleqn,12pt,twoside]{article}
\usepackage{espcrc1}
\usepackage{epsf}

\newcommand{\half}{\textstyle \frac{1}{2}}

\newcommand{\mth}{m_\Theta}

\newcommand{\eth}{\eta_\Theta}

\newcommand{\AmS}{{\protect\the\textfont2
  A\kern-.1667em\lower.5ex\hbox{M}\kern-.125emS}}

\title{\begin{flushright} \small DESY-05-005 \\ PC 077.1204
  \end{flushright} 
Exclusive production of pentaquarks in the scaling
regime\footnote{Talk given at the 10th International Conference on the
Structure of Baryons (Baryons 2004), Palaiseau, France, 25--29 October
2004.  To appear in the Proceedings.}}

\author{M. Diehl\address{Deutsches Elektronen-Synchroton DESY, 22603
        Hamburg, Germany},
        B.~Pire\address{CPhT, \'Ecole Polytechnique, 
        91128 Palaiseau, France}\thanks{UMR 7644 du CNRS}         
        L.~Szymanowski\address{Soltan Institute for Nuclear Studies,
        Ho\.{z}a 69, 00-681 Warsaw, Poland and \\ 
        Universit\'e de Li\`ege, 4000 Li\`ege, Belgium}
        }
       
\begin{document}

\maketitle

\begin{abstract}
We investigate two exclusive reactions with a $\Theta^+$ pentaquark in
the final state: electroproduction of a $K$ meson on the nucleon, and
$K^+$ scattering on a neutron target producing a lepton pair.  These
reactions offer unique opportunities to investigate the structure of
pentaquark baryons at parton level.  We discuss the generalized parton
distributions for the $N \to \Theta^+$ transition and give the leading
order amplitude for these processes in the Bjorken regime.
\end{abstract}

\vfill

\section{Motivation}

Several experiments have reported evidence for the existence of a
narrow baryon resonance $\Theta^+$ with strangeness $S=+1$, whose
minimal quark content is $uudd\bar{s}$ \cite{Nakano}.  Triggered by
the prediction of its mass and width in
\cite{Diakonov:1997mm,Praszalowicz:2003ik}, the observation of this
hadron promises to shed new light on our picture of baryons in QCD.  A
fundamental question is how the structure of baryons manifests itself
in terms of the basic degrees of freedom in QCD, at the level of
partons.  This structure at short distances can be probed in hard
exclusive scattering processes, where it is encoded in generalized
parton distributions \cite{Muller:1994fv,Reviews}.  In Ref.~\cite{DPS}
we introduced the transition GPDs from the nucleon to the $\Theta^+$
and investigated electroproduction processes where they could be
measured, hopefully already in existing experiments at DESY and
Jefferson Lab.  The crossed channel, where a timelike virtual photon
is produced together with the pentaquark, might be studied at intense
kaon beam facilities presently under discussion \cite{Nagamiya}.

\newpage


\section{Processes}
\label{sec:channels}

\begin{figure}
\begin{center}
    \epsfxsize=0.95\textwidth
    \epsffile{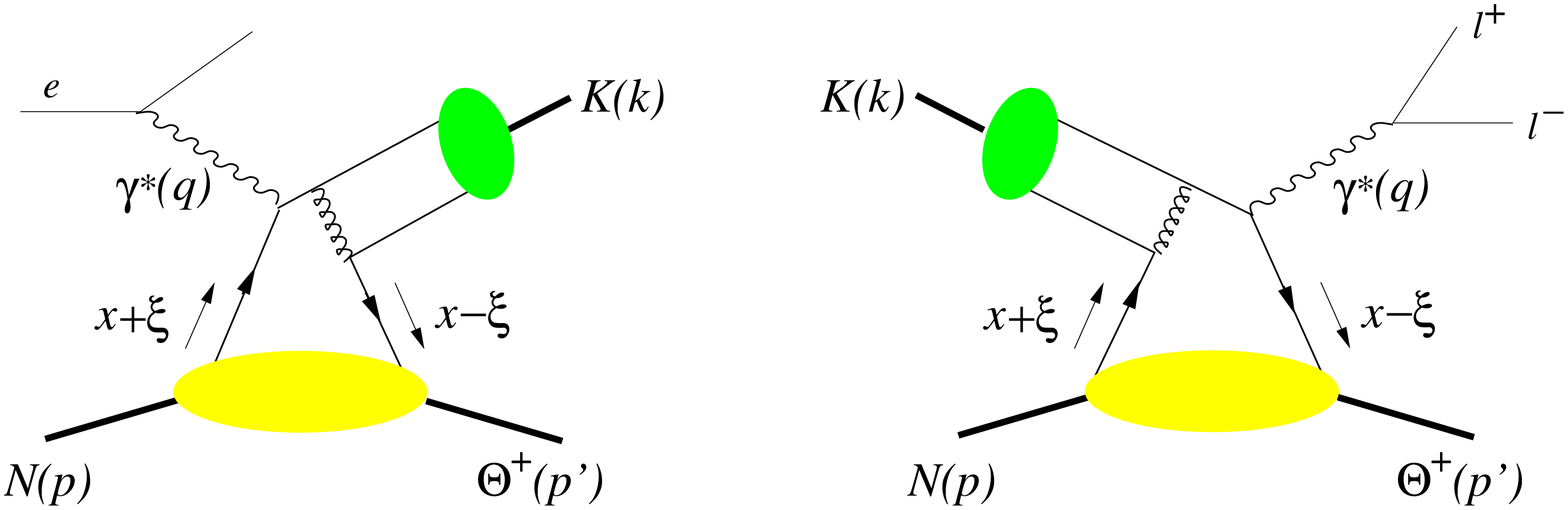}
\caption{\label{fig:proc} Example diagrams at leading order in
$\alpha_s$ for pentaquark electroproduction (left) and its timelike
counterpart (right) in the scaling limit.  In both cases three further
diagrams are obtained by attaching the photon to the quark lines in
all possible ways.  The plus-momentum fractions $x$ and $\xi$ refer to
the average momentum $P=\frac{1}{2}(p+p')$ of nucleon and pentaquark.}
\end{center}
\end{figure}

We consider the processes
\begin{eqnarray}
  \label{proc-p}
e p &\to & e \bar{K}^0 \, \Theta^+ , \qquad \qquad
e n\to e K^- \, \Theta^+ , \\
  \label{K-n}
K^+ n &\to & \ell^+\,\ell^- \, \Theta^+ ,
\end{eqnarray}
where $\ell$ is an electron or muon and where the $\Theta^+$
subsequently decays into $K^0 p$ or $K^+ n$.  The case of $K^*$
electroproduction was also discussed in Ref.~\cite{DPS}.  We are
interested in the Bjorken limit, which for the electroproduction
processes (\ref{proc-p}) is given by large $Q^2= -q^2$ at fixed
$t=(p-p')^2$ and fixed scaling variable $x_B = Q^2 /(2 pq)$.  For the
crossed process (\ref{K-n}) we take the limit of large $Q^2= q^2$ at
fixed $t=(p-p')^2$ and fixed $\tau = {Q^2}/(2 pk)$.  Four-momenta are
specified in Fig.~\ref{fig:proc}.

According to the factorization theorem for meson production
\cite{Collins:1997fb}, the Bjorken limit implies factorization of the
exclusive amplitudes into a perturbatively calculable subprocess at
quark level, the distribution amplitude of the meson, and a
generalized parton distribution (GPD) describing the transition from
the nucleon to the $\Theta^+$ (see Fig.~\ref{fig:proc}).  The
arguments for factorization in \cite{Collins:1997fb} do not rely on
the photon being spacelike and can be extended to the case $K^- n \to
\gamma^*\, \Theta^+$ with the same nonperturbative input.  The
dominant polarization of the spacelike or timelike photon is then
longitudinal, and the corresponding cross sections have a
characteristic scaling behavior as we will see below.


\section{The transition GPDs and their physics}
\label{sec:gpds}

To define the transition GPDs we introduce light-cone coordinates
$v^\pm = (v^0 \pm v^3) /\sqrt{2}$ and transverse components $v_T =
(v^1, v^2)$ for any four-vector $v$.  The skewness variable $\xi =
(p-p')^+ /(p+p')^+$ describes the loss of plus-momentum of the
incident nucleon and is connected with $x_B$ or $\tau$ by $\xi \approx
{x_B}/{(2-x_B)}$ or $\xi \approx {\tau}/{(2-\tau)}$¥ in the Bjorken
limit.  We assume that the $\Theta^+$ has spin $J=\half$, isospin
$I=0$ and intrinsic parity either $\eth = 1$ or $\eth = -1$.  The
hadronic matrix elements needed at leading-twist accuracy are
\begin{eqnarray}
  \label{matrix-elements}
F_V &=&
\frac{1}{2} \int \frac{d z^-}{2\pi}\, e^{ix P^+ z^-}
  \langle \Theta^+|\, \bar{d}(-\half z)\, \gamma^+ s(\half z) 
  \,|p \rangle \Big|_{z^+=0,\, {z}_T=0} \; ,
\nonumber \\
F_A &=&
\frac{1}{2} \int \frac{d z^-}{2\pi}\, e^{ix P^+ z^-}
  \langle \Theta^+|\, 
     \bar{d}(-\half z)\, \gamma^+ \gamma_5\, s(\half z)
  \,|p \rangle \Big|_{z^+=0,\, {z}_T=0}
\end{eqnarray}
with $P = \half (p+p')$.  The corresponding $p\to \Theta^+$ transition
GPDs are defined by
\begin{eqnarray}
  \label{gpd-pos}
F_V &=& \frac{1}{2P^+} \left[
  H(x,\xi,t)\, \bar{u}(p') \gamma^+ u(p) +
  E(x,\xi,t)\, \bar{u}(p') 
                 \frac{i \sigma^{+\alpha} (p'-p)_\alpha}{\mth+m_N} u(p)
  \, \right] ,
\nonumber \\
F_A &=& \frac{1}{2P^+} \left[
  \tilde{H}(x,\xi,t)\, \bar{u}(p') \gamma^+ \gamma_5 u(p) +
  \tilde{E}(x,\xi,t)\, \bar{u}(p') 
\frac{\gamma_5\, (p'-p)^+}{\mth+m_N} u(p)
  \, \right]
\end{eqnarray}
for $\eth = 1$ and
\begin{eqnarray}
  \label{gpd-neg}
F_V &=& \frac{1}{2P^+} \left[
  \tilde{H}(x,\xi,t)\, \bar{u}(p') \gamma^+ \gamma_5 u(p) +
  \tilde{E}(x,\xi,t)\, \bar{u}(p') 
\frac{\gamma_5\, (p'-p)^+}{\mth+m_N} u(p)
  \, \right] ,
\nonumber \\
F_A &=& \frac{1}{2P^+} \left[
  H(x,\xi,t)\, \bar{u}(p') \gamma^+ u(p) +
  E(x,\xi,t)\, \bar{u}(p') 
        \frac{i \sigma^{+\alpha} (p'-p)_\alpha}{\mth+m_N} u(p)
  \, \right]
\end{eqnarray}
for $\eth = -1$, where $m_N$ and $\mth$ denote the mass of the nucleon
and the $\Theta^+$, respectively.  Due to isospin symmetry the GPDs
for the transition $n\to \Theta^+$ are equal to those for $p\to
\Theta^+$ up to a global sign.

As shown in \cite{Burkardt:2000za}, GPDs contain information about the
spatial structure of hadrons.  A Fourier transform of their dependence
on $t$ tells us about the transverse size of the hadrons in question.
For instance, for $\xi<|x|<1$ the transverse positions of all partons
must match in the proton and the $\Theta^+$, including the quark or
antiquark taking part in the hard scattering.  Thus, the $p\to
\Theta^+$ transition GPDs probe the partonic structure of the
$\Theta^+$ by requiring its wave function to overlap with the wave
function of appropriate configurations in the nucleon.


\section{Scattering amplitude and cross section}
\label{sec:scatter}

At leading order in $1/Q$ and in $\alpha_s$, the factorization theorem
gives scattering amplitudes
\begin{eqnarray}
  \label{amp-p}
\mathcal{A}(\gamma^* p\to \bar{K}^0\, \Theta^+) &=&
i e\, \frac{8\pi\alpha_s}{27}\, \frac{f_K}{Q}\, \Bigg[
I_K \int_{-1}^1 \frac{dx}{\xi-x-i\epsilon}\, 
\Big( F_A(x,\xi,t) - F_A(-x,\xi,t) \Big)
\\
 && \hspace{4.2em} {}+
J_K \int_{-1}^1 \frac{dx}{\xi-x-i\epsilon}\,  
\Big( F_A(x,\xi,t) + F_A(-x,\xi,t) \Big)
\, \Bigg] ,
\nonumber \\
\mathcal{A}(\gamma^* n\to K^-\, \Theta^+) &=&
-i e\, \frac{8\pi\alpha_s}{27}\, \frac{f_K}{Q}\, \Bigg[
I_K \int_{-1}^1 \frac{dx}{\xi-x-i\epsilon}\, 
\Big( F_A(x,\xi,t) + 2 F_A(-x,\xi,t) \Big)
\nonumber \\
 && \hspace{5em} {}+
J_K \int_{-1}^1 \frac{dx}{\xi-x-i\epsilon}\,  
\Big( F_A(x,\xi,t) - 2 F_A(-x,\xi,t) \Big)
\, \Bigg] 
\nonumber 
\end{eqnarray}
for longitudinal polarization of the photon, independently of the
parity of the $\Theta^+$.  Here $I$ and $J$ are definite integrals
over the twist-two distribution amplitude of the kaon, given in
\cite{DPS}.  At fixed $\xi$ and $t$ the amplitude scales like $1/Q$,
up to logarithmic corrections from the running of $\alpha_s$ and from
the scale evolution of the GPDs and distribution amplitudes.  The
corresponding $ep$ or $en$ cross section is obtained from
\begin{eqnarray}
  \label{X-section}
\frac{d\sigma}{dQ^2\, dt\, dy} &=& 
\frac{\alpha_{\mathit{em}}}{32\pi^2}\, \frac{1-y}{y}\,
\frac{x_B^2}{Q^6}\, \sum_{\lambda',\lambda}
\Big| \mathcal{A}_{\lambda',\lambda} \Big|^2  ,
\end{eqnarray}
where $\lambda$ ($\lambda'$) denotes the helicity of the incoming
(outgoing) baryon and where $y$ is the usual inelasticity parameter
introduced in deep inelastic scattering.  Realistic cross section
estimates based on these expressions are unfortunately not possible at
present, due to our ignorance of the $p\to \Theta^+$ transition GPDs.

The amplitude for the crossed process (\ref{K-n}) is obtained in a
straightforward way \cite{BDP}.  At equal $Q^2$, $t$ and $\xi$ one has
\begin{eqnarray}
\mathcal{A}_{\lambda',\lambda}(K^+ n \to \gamma^* \Theta^+)
  &=& \Big[ \mathcal{A}_{\lambda',\lambda}(\gamma^* n\to K^- \Theta^+)
      \Big]^*
\label{relation}
\end{eqnarray}
at leading power in $1/Q$ and in $\alpha_S$, up to an irrelevant
global phase and provided that the phase convention for helicity
spinors gives real valued matrix elements in (\ref{gpd-pos}) and
(\ref{gpd-neg}).  We anticipate that the relation (\ref{relation}) no
longer holds at the level of corrections in $\alpha_S$ or in $1/Q$.
The cross section for the overall process $K^+ n \to \ell^+\ell^-\,
\Theta^+$ is obtained from
\begin{equation}
\frac{d\sigma}{dQ^2\, dt\, d(\cos\theta)}
= \frac{\alpha_{\mathrm{em}}}{128 \pi^2} \sin^2\theta\,
  \frac{\tau^2}{Q^6}\,
  \sum_{\lambda',\lambda} 
  \Big| \mathcal{A}_{\lambda',\lambda} \Big|^2 ,
\label{cross-section}
\end{equation}
where $\theta$ is the polar decay angle of the photon in its rest
frame (cf.\ Fig.~5 of \cite{Berger:2001}).  The $\sin^2\theta$
behavior in (\ref{cross-section}) reflects the purely longitudinal
$\gamma^*$ polarization.  Note that in this process the $\Theta^+$ is
produced in the very clean environment of a final state with only the
resonance and a high-mass lepton pair.


\section*{Acknowledgments}  
This work is supported by the Joint Research Activity "Generalised
Parton Distributions" in Integrated Infrastructure Initiative ``Hadron
Physics'' of the European Union, contract No.~RII3-CT-2004-506078 
and by the  Polish Grant 1 P03B 028 28.
L.Sz.\ is a Visiting Fellow of the Fonds National pour la Recherche
Scientifique (Belgium).  The work of B.P. and L. Sz.\ is partially
supported by the French-Polish scientific agreement Polonium.


\end{document}